\documentstyle[aps,12pt]{revtex}

\begin{document}
\author{J.P. Krisch and E.N. Glass\thanks{%
Permanent address: Physics Department, University of Windsor, Ontario N9B
3P4, Canada}}
\title{Adding\ Twist to Anisotropic Fluids}
\address{Department of Physics, University of Michigan, Ann Arbor, Michgan 48109}
\date{November 13, 2001 \ \ \ preprint MCTP-01-62}
\maketitle

\begin{abstract}
We present a solution generating technique for anisotropic fluids which
preserves specific Killing symmetries. Anisotropic matter distributions that
can be used with the one parameter Ehlers-Geroch transform are discussed.
Example spacetimes that support the appropriate anisotropic stress-energy
are found and the transformation applied. The 3+1 black string solution is
one of the spacetimes with the appropriate matter distribution. Use of the
transform with a black string seed is discussed.\newline
\newpage
\end{abstract}

\section{Introduction}

Physically relevant solution generating techniques were developed in the
1960s and 1970s. Ehlers \cite{Ehl62}, Harrison \cite{Har68}, and Geroch \cite
{Ger71},\cite{Ger72} showed that a projective transform on the norm and
twist ($\lambda ,\omega )$, of a Killing vector will generate a Killing
vector with norm and twist ($\lambda ^{\prime },\omega ^{\prime }$).
Starting with a vacuum spacetime and a twist-free Killing vector, their
method adds twist. For example, their method applied to the vacuum
Schwarzschild metric generates a NUT metric. The generating method can be
applied to any vacuum spacetime with a Killing vector and has been
generalized to the\ Einstein-Maxwell spaces \cite{Kin73} and to some matter
spacetimes \cite{Bel79}. The extension to matter metrics is restrictive;
Stephani \cite{Ste88}\ has shown that the only two equations of state that
can be treated within this formalism are
\begin{eqnarray}
\rho &=&P, \\
\rho +3P &=&0.  \nonumber
\end{eqnarray}
Raca and\ Zsigrai \cite{RZ96} have also considered solution generating on
fluids with this equation of state. The result clearly applies to fluids
with isotropic stress. A close examination of the method used to generate
the allowed matter distributions shows that it can be generalized to fluids
with anisotropic stresses.

There has been increasing interest in general relativistic systems with
anisotropic stress. Herrera and Santos \cite{HS97} have reviewed some of the
possible causes and the related general relativistic solutions. Anisotropic
fluid spheres have been a useful model for discussing anisotropy since the
early work of Bowers and Liang \cite{BL74} on anisotropic fluid spheres.
More recently Corchero \cite{Cor98} has discussed a post-Newtonian
approximation to anisotropic fluid spheres. Das et al \cite{DTA+98} and Das
and Kloster \cite{DK00} have investigated the spherically symmetric collapse
of anisotropic fluid objects into a black hole. Hernandez, Nunez, and
Percoco \cite{HNP99} have treated anisotropy and non-local equations of
state in radiating spheres. Conformally flat anisotropic spheres were
examined by Herrera et al \cite{HDO+01}. Glass and Krisch \cite{GK98} have
discussed diffusion induced anisotropies in a Vaidya atmosphere. Following
up the recent interest in dimensional effects, Harko and Mak \cite{HM00}
discussed charged anisotropic fluid spheres in D-dimensions. Anisotropy in
cosmological solutions has also been studied by McManus and Coley \cite{MC94}%
, vandenHoogen and Coley \cite{vHC95},\ Giovannini \cite{Gio99} and
Rainsford \cite{Rai01}. Relative motion as a source of anisotropy in
multi-fluid systems was suggested long ago by Jeans \cite{Jea22}. Letelier
\cite{Let80} has shown that two perfect fluids in relative motion can be
described as a system with anisotropic pressures and has given the standard
two-fluid stress energy form.

Physically, anisotropy is finding increasing application in systems at the
density extremes of very compact objects and very diffuse mass clusters.
This increasing interest in fluid solutions with anisotropic pressures
suggests it would be of interest to extend the solution generating technique
to spacetimes that have anisotropic fluid content. While it is relatively
simple to create anisotropic fluid solutions by changing the functional
dependence of metric potentials, the generating technique here will preserve
specific Killing symmetries while obtaining new anisotropic solutions.

In the next section, we briefly review the formalism that leads to the
isotropic pressure restrictions on the standard solution generating
technique. The extension to anisotropic stresses follows directly from this.
Isotropic seed spacetimes are discussed in section III. Some examples of
spacetimes containing the allowed anisotropies are in section IV. Killing
symmetries for the possible anisotropic spaces are covered in section V, and
section VI has a discussion of the effect of the Geroch transformation on
the fluid description in these spaces.

\section{Formalism}

\subsection{The Killing Description}

Let ($M,g_{ab}$) be a solution of the Einstein field equations with energy
density $\rho $ and isotropic pressure $P$. Assume that $g_{ab}$ has a
Killing vector $\xi ^{a}$ with norm $\lambda $ and twist $\omega ^{a}$ where
\begin{eqnarray}
\lambda &:&=\xi ^{a}\xi _{a},  \label{norm-twist} \\
{\omega }^{a} &:&={\epsilon ^{abcd}}{\xi _{b}}{\nabla _{c}}{\xi }_{d}.
\nonumber
\end{eqnarray}
The induced metric on the 3-dimensional space of vectors orthogonal to the
Killing vector is
\begin{equation}
h_{ab}=g_{ab}-\xi _{a}\xi _{b}/\lambda .  \label{h-met}
\end{equation}
The vacuum field equations can be written in terms of $\lambda ,$\ $\omega $
and $\gamma _{ab}$
\begin{eqnarray}
\gamma _{ab} &=&\left| \lambda \right| \ h_{ab},  \label{gam-met} \\
\omega _{a} &=&D_{a}\omega ,  \nonumber
\end{eqnarray}
where $\gamma _{ab}$ is conformally related to $h_{ab}$ \cite{Ger71}, $%
\omega $ is the scalar potential associated with the Killing twist, and $%
D_{a}$ is the covariant derivative for metric $\gamma _{ab}.$

\subsection{Vacuum Spacetimes}

Consider the action written in the conformal 3-space of $\gamma _{ab}$%
\begin{equation}
I=\int d^{3}x\sqrt{\gamma }\left[ {\cal R-}\frac{1}{2\lambda ^{2}}%
(D_{a}\lambda D^{a}\lambda +D_{a}\omega D^{a}\omega )\right] .
\end{equation}
A projective transform of the complex potential $\tau =\omega +i\lambda $
can be written
\begin{equation}
\tau ^{\prime }=\frac{\tau \cos (\delta )+\sin (\delta )}{\cos (\delta
)-\tau \sin (\delta )}  \label{tau-prime}
\end{equation}
where $\delta $ is a transformation parameter. The action is invariant under
this transformation, so we have added twist to the Killing vector with $%
\gamma _{ab\text{ }}$ unchanged and can generate the new 3+1 spacetime, $%
g_{ab}^{\prime },$ from the formalism. The development of the transformation
method is described in references \cite{Ger71},\cite{Ger72}. We briefly
review the transformation method for twist-free Killing vectors.

\subsection{Generating the new spacetime from a twist-free seed}

The generation method as described by Geroch \cite{Ger71},\cite{Ger72} for
vacuum will be the same in the matter spacetimes. Starting with metric $%
g_{ab}$, define forms $\alpha _{a},$ $\beta _{a}$, and $\eta _{a}$ based on
a seed Killing vector, $\xi _{a}$, with norm $\lambda $ and transformation
parameter $\delta $:
\begin{mathletters}
\label{new-norm-twist}
\begin{eqnarray}
\beta _{a} &=&\xi _{a}(\lambda -\lambda ^{-1}), \\
2\alpha _{\lbrack b,a]} &=&\varepsilon _{abcd}\nabla ^{c}\xi ^{d}, \\
\eta _{a} &=&\lambda ^{-1}\xi _{a}+\alpha _{a}\sin (2\delta ).
\end{eqnarray}
The new metric is then given by
\end{mathletters}
\begin{equation}
g_{ab}^{\prime }=F(g_{ab}-\lambda ^{-1}\xi _{a}\xi _{b})+(\lambda /F)\eta
_{a}\eta _{b}  \label{g-prime-met}
\end{equation}
with
\begin{equation}
F=\cos ^{2}(\delta )+\lambda ^{2}\sin ^{2}(\delta ).  \label{f}
\end{equation}
The norm of the Killing vector becomes
\begin{equation}
\lambda ^{\prime }=\lambda /F.  \label{lambda-prime}
\end{equation}
The scalar twist potential that has been added to the Killing vector is
\begin{equation}
\omega ^{\prime }=\frac{\sin (\delta )\cos (\delta )(1-\lambda ^{2})}{F}.
\label{omega-prime}
\end{equation}
These are the equations that will generate the new spacetime from static
seed metrics.

\section{Isotropic Matter Descriptions}

\subsection{Isotropic Model 1}

To find the isotropic matter spacetimes that can be used with this method,
consider the action \cite{Gar01}
\begin{equation}
I=\int d^{3}x\sqrt{\gamma }\left[ {\cal R-}\frac{D_{a}\lambda D^{a}\lambda }{%
2\lambda ^{2}}+\Psi \right]
\end{equation}
with $\Psi $ a specified function. Stephani \cite{Ste88} has shown that the
Ricci tensor in 4-dimensions is
\begin{equation}
R_{ab}=-\Psi \left| \lambda \right| (g_{ab}-\xi _{a}\xi _{b}/\lambda )
\end{equation}
and clearly the action will be invariant under the projective transform on $%
\tau .$ In the isotropic case, this Ricci tensor is associated with a
perfect fluid for $\xi ^{a}$ timelike. The fluid has an equation of state $%
\rho +3P=0,$ with $8\pi \rho =3\Psi \lambda /2.$

\subsection{Isotropic Model 2}

A second action that one can consider is \cite{Gar01}
\begin{equation}
I=\int d^{3}x\sqrt{\gamma }\left[ {\cal R-}\frac{D_{a}\lambda D^{a}\lambda }{%
2\lambda ^{2}}-s_{a}s^{a}\right]
\end{equation}
with
\begin{equation}
s_{a}\xi ^{a}=0.
\end{equation}
The 4-dimensional Ricci tensor for this action is
\begin{equation}
R_{ab}=s_{a}s_{b}
\end{equation}
which, in the isotropic case, describes a perfect fluid if the Killing
vector is spacelike. The fluid has density and pressure
\begin{eqnarray}
8\pi \rho &=&-s^{a}s_{a}/2, \\
P &=&\rho .  \nonumber
\end{eqnarray}
Solution generating with both of these isotropic forms has been examined by
Garfinkle, Glass, and\ Krisch \cite{GGK97}.

\section{Anisotropic Matter Description}

The form of the Ricci tensor will be the same for fluids with non-isotropic
stress. The stress-energy content can be written as
\begin{equation}
(1/8\pi )T_{ab}=\rho
u_{a}u_{b}+p_{1}e_{a}^{(1)}e_{b}^{(1)}+p_{2}e_{a}^{(2)}e_{b}^{(2)}+p_{3}e_{a}^{(3)}e_{b}^{(3)}
\end{equation}
where $(u^{a},e_{a}^{(i)},\ \ i=1,2,3)$ is a convenient orthogonal tetrad.
The associated Ricci tensor is
\begin{eqnarray}
R_{ab} &=&T_{ab}-(T/2)g_{ab} \\
(1/8\pi )R_{ab} &=&\rho
u_{a}u_{b}+p_{1}e_{a}^{(1)}e_{b}^{(1)}+p_{2}e_{a}^{(2)}e_{b}^{(2)}+p_{3}e_{a}^{(3)}e_{b}^{(3)}+(%
\frac{\rho -p_{\Sigma }}{2})g_{ab}
\end{eqnarray}
where $p_{\Sigma }=p_{1}+p_{2}+p_{3}$.

\subsection{Anisotropic Model 1}

Consider Ricci tensor $R_{ab}=-\Psi \left| \lambda \right| (g_{ab}-\xi
_{a}\xi _{b}/\lambda )$. There are two Killing vector possibilities: $\xi
_{a}=\lambda u_{a}$ and $\xi _{a}=\lambda e_{a}^{(1)},$ where we have chosen
$e_{a}^{(1)}$ for convenience. The timelike Killing vector will generate
isotropic stress. Consider the spacelike vector $\xi _{a}=\lambda
e_{a}^{(1)}.$ We have
\begin{equation}
-(1/8\pi )\Psi \left| \lambda \right| (g_{ab}-e_{a}^{(1)}e_{b}^{(1)})=\rho
u_{a}u_{b}+p_{1}e_{a}^{(1)}e_{b}^{(1)}+p_{2}e_{a}^{(2)}e_{b}^{(2)}+p_{3}e_{a}^{(3)}e_{b}^{(3)}+(%
\frac{\rho -p_{\Sigma }}{2})g_{ab}.
\end{equation}
Multiplying by $e_{a}^{(1)}$ we have
\begin{equation}
p_{1}+\rho =p_{2}+p_{3}
\end{equation}
Contracting with $u_{a}$ we find$\ $%
\begin{equation}
(1/8\pi )2\Psi \left| \lambda \right| =\rho +p_{\Sigma }.
\end{equation}
The other spatial contractions give
\begin{eqnarray}
-(1/8\pi )2\Psi \left| \lambda \right|  &=&p_{2}-p_{1}-p_{3}+\rho \  \\
-(1/8\pi )2\Psi \left| \lambda \right|  &=&p_{3}-p_{1}-p_{2}+\rho   \nonumber
\end{eqnarray}
Comparing, we must have $p_{2}=p_{3}$, and
\begin{equation}
-(1/8\pi )2\Psi \left| \lambda \right| =-p_{1}+\rho
\end{equation}
Combining, we find an anisotropic fluid with density and stress
\begin{eqnarray}
8\pi \rho  &=&-\Psi \lambda /2 \\
8\pi p_{1} &=&3\Psi \lambda /2  \nonumber \\
p_{2} &=&p_{3}=-\rho .  \nonumber
\end{eqnarray}

\subsection{Anisotropic Model 2}

Consider Ricci tensor $R_{ab}=s_{a}s_{b}$. This Ricci tensor model requires
that the vector $s^{a}$ be orthogonal to the Killing vector. Again there are
two choices for the Killing vector, and in this model it is the timelike
Killing vector that generates the anisotropic stress energy. Consider $\xi
_{a}=\lambda u_{a}.$ Since $s_{a}$ is orthogonal to the Killing vector, $%
s_{a}$ is spacelike. Choose function $\Phi $ and $s_{a}=\Phi e_{a}^{(1)}$.
\begin{equation}
\frac{s_{a}s_{b}}{8\pi }=\frac{\Phi ^{2}e_{a}^{(1)}e_{b}^{(1)}}{8\pi }=\rho
u_{a}u_{b}+p_{1}e_{a}^{(1)}e_{b}^{(1)}+p_{2}e_{a}^{(2)}e_{b}^{(2)}+p_{3}e_{a}^{(3)}e_{b}^{(3)}+(%
\frac{\rho -p_{\Sigma }}{2})g_{ab}
\end{equation}
Following the same method used in the previous section, we obtain
\begin{eqnarray}
8\pi \rho  &=&8\pi p_{1}=\Phi ^{2}/2 \\
8\pi p_{2} &=&8\pi p_{3}=-\Phi ^{2}/2.  \nonumber
\end{eqnarray}
The indices can be relabeled to describe $s_{a}$ lying along $e_{a}^{(2)}$
or $e_{a}^{(3)}$.

\section{Spacetimes for the anisotropic matter distributions}

In the previous section we examined two anisotropic models. One has a
timelike Killing vector and one a spacelike Killing vector. We now find
examples of spacetimes that could contain the anisotropic matter
distribution.

\subsection{Timelike Killing vector}

For density and pressures $\rho =p_{1}=-p_{2}=-p_{3}\ $consider the metric
with function $\chi (z)$:
\begin{equation}
ds^{2}=-e^{2n\chi }dt^{2}+dz^{2}+e^{2\chi }(dr^{2}+r^{2}d\varphi ^{2}).
\end{equation}
The field equations are
\begin{eqnarray}
8\pi \rho &=&-2\chi _{,zz}-3\chi _{,z}^{2} \\
8\pi p_{z} &=&\chi _{,z}^{2}(2n+1)  \nonumber \\
8\pi p_{k} &=&(n+1)\chi _{,zz}+\chi _{,z}^{2}(n^{2}+n+1)  \nonumber
\end{eqnarray}
with $p_{k}$ labelling both $p_{r}$ and $p_{\varphi }.$ Enforcing the stress
relations, one finds the solution
\[
e^{(n+2)\chi }=az+b
\]
with fluid $\rho =p_{z}=-p_{k}$%
\begin{equation}
8\pi \rho =\frac{(2n+1)a^{2}}{(n+2)^{2}(az+b)^{2}}
\end{equation}
and metric
\begin{equation}
ds^{2}=-(az+b)^{2n/(n+2)}dt^{2}+dz^{2}+(az+b)^{2/(n+2)}(dr^{2}+r^{2}d\varphi
^{2}).
\end{equation}
For $n=1$, this spacetime is conformally flat.

\subsection{Spacelike Killing vector}

A simple spacetime whose fluid content has the necessary anisotropic
structure is the conformally flat spacetime with metric
\begin{equation}
ds^{2}=e^{2az}(-dt^{2}+dr^{2}+r^{2}d\varphi ^{2}+dz^{2}).
\end{equation}
The fluid parameters are easily shown to be
\begin{eqnarray}
8\pi \rho  &=&-a^{2}e^{-2az} \\
8\pi p_{z} &=&3a^{2}e^{-2az}  \nonumber \\
8\pi p_{k} &=&a^{2}e^{-2az}  \nonumber
\end{eqnarray}
where $p_{k}$ are the radial and $\varphi -$stresses. The negative density
does not readily lend itself to a physical description. An interesting
spacetime that also has the appropriate anisotropic stress relations is the
simple lift of the 2+1 BTZ black hole spacetime \cite{BTZ92} describing an
infinite black string \cite{Kal93},\cite{CM95}:
\begin{equation}
ds^{2}=-(-m+\Lambda _{3}r^{2})dt^{2}+\frac{dr^{2}}{-m+\Lambda _{3}r^{2}}%
+r^{2}d\varphi ^{2}+dz^{2}.
\end{equation}
In 2+1 there is a stress energy
\[
T_{ii}=\Lambda _{3}g_{ii}
\]
with $\Lambda _{3}$ the 2+1 cosmological constant. When the $z$ coordinate
is added the fluid content is
\begin{eqnarray}
8\pi \rho  &=&-\Lambda _{3} \\
8\pi p_{r} &=&8\pi p_{\varphi }=\Lambda _{3}  \nonumber \\
8\pi p_{z} &=&3\Lambda _{3}  \nonumber
\end{eqnarray}
which has the required anisotropic stress-energy structure. The relation
between the 2+1 BTZ solution and the 3+1 black string has been studied by
Lemos and Zanchin \cite{LZ96}. The negative density in this case can be
physically motivated from the cosmological constant. It will be of interest
to apply the Geroch transform to this 3+1 BTZ lift and then project back
down to 2+1 to examine the effects on the cosmological fluid.

\section{Applying the Geroch Transformation}

In this section we will use the Geroch formalism described by Eqs.(\ref
{new-norm-twist})-(\ref{lambda-prime}) to add twist to the Killing vectors
of our example spacetimes. The new spacetime will be generated and the
effect of the transformation on the fluid parameters examined.

\subsection{Timelike Killing vector}

The spacetime with a timelike Killing vector that we found had metric
\[
ds^{2}=-(az+b)^{2n/(n+2)}dt^{2}+dz^{2}+(az+b)^{2/(n+2)}(dr^{2}+r^{2}d\varphi
^{2}).
\]
The Geroch transform process can be applied to this spacetime adding twist
to the timelike Killing vector and vorticity to the fluid. From Eqs.(\ref
{new-norm-twist})-(\ref{f}) we have
\begin{eqnarray}
\xi _{a}^{(t)}\xi _{(t)}^{a} &=&\lambda =-(az+b)^{2n/(n+2)} \\
F &=&\cos ^{2}(\delta )+\lambda ^{2}\sin ^{2}(\delta ) \\
\alpha _{\varphi ,r}-\alpha _{r,\varphi } &=&-\frac{2nar}{n+2}\eta
_{tr\varphi z} \\
\alpha _{\varphi } &=&-\frac{nar^{2}}{n+2}+\alpha _{0}
\end{eqnarray}
where $\eta _{tr\varphi z}=1$. The new metric is
\begin{equation}
ds^{2}=-\frac{(az+b)^{2n/(n+2)}}{F}[dt+\sin (2\delta )\alpha _{\varphi
}d\varphi ]^{2}+Fdz^{2}+F(az+b)^{2/(n+2)}(dr^{2}+r^{2}d\varphi ^{2}).
\end{equation}
The fluid parameters in this spacetimes are the original parameters scaled
by $F$:
\begin{eqnarray}
\rho ^{\prime } &=&\rho /F \\
p_{i}^{\prime } &=&p_{i}/F.  \nonumber
\end{eqnarray}
The fluid has acquired vorticity along the z-axis
\begin{equation}
\omega ^{(z)}=\frac{2na\sin (2\delta )}{(n+2)F^{2}}\ (az+b)^{(n-2)/(n+2)}.
\end{equation}
The projective transform on the Killing parameters that generates the new
spacetime has two fixed points. For the case where the initial space is
twist free, the fixed points of the projective transform are $\lambda =\pm
1. $ For this example, $F=1$ at the fixed points, and the fluid parameters
are the same in both the seed and transformed spacetimes.

\subsection{Spacelike Killing vector}

In this example, we wish to add twist to a spacelike Killing vector. We will
consider the lift of the BTZ metric as the seed spacetime. The seed metric
is
\begin{equation}
ds^{2}=-(-m+\Lambda _{3}r^{2})dt^{2}+\frac{dr^{2}}{-m+\Lambda _{3}r^{2}}%
+r^{2}d\varphi ^{2}+dz^{2}.
\end{equation}
There are two Killing vector choices: $\xi _{a}^{(\varphi )}$ or $\xi
_{a}^{(z)}.$ We will work with the $\varphi $-Killing vector and assume that
the metric remains independent of the z-coordinate. From Eq.(\ref
{new-norm-twist}) we have
\begin{eqnarray}
\xi _{a}^{(\varphi )}\xi _{(\varphi )}^{a} &=&\lambda =r^{2} \\
\alpha _{t,z}-\alpha _{z,t} &=&2\eta _{tzr\varphi }(-m+\Lambda _{3}r^{2})
\end{eqnarray}
Calculating $\alpha _{a}$ and $\eta _{a}$ we have
\begin{equation}
\alpha _{z}=2(m-\Lambda _{3}r^{2})t+\alpha _{0},
\end{equation}
\begin{equation}
\eta _{a}=\xi _{a}^{(\varphi )}/\lambda +\sin (2\delta )\alpha _{a},
\end{equation}
with $F=\cos ^{2}$($\delta $)$+r^{4}$sin$^{2}$($\delta $). The new 3+1
metric is
\begin{equation}
ds_{\ }^{2}=F[-(-m+\Lambda _{3}r^{2})dt^{2}+\frac{dr^{2}}{-m+\Lambda
_{3}r^{2}}+dz^{2}]+\frac{r^{2}}{F}[d\varphi +\alpha _{z}\sin (2\delta
)dz]^{2}.  \label{new3+1-met}
\end{equation}
This black string solution is an anisotropic member of the cylindrical black
hole solution family discussed by Lemos \cite{Le95}.

Using $\delta =\pi /2$, it is easy to see the effect of the transform on the
fluid parameters. For this value, the new 3+1 spacetime becomes$\ $%
\begin{equation}
ds^{2}=r^{4}[-(-m+\Lambda _{3}r^{2})dt^{2}+\frac{dr^{2}}{-m+\Lambda _{3}r^{2}%
}+dz^{2}]+\frac{1}{r^{2}}d\varphi ^{2}
\end{equation}
with a fluid content
\begin{eqnarray}
8\pi \rho  &=&-\Lambda _{3}/r^{4}, \\
8\pi p_{r} &=&\Lambda _{3}/r^{4},  \nonumber \\
8\pi p_{\varphi } &=&9\Lambda _{3}/r^{4},  \nonumber \\
8\pi p_{z} &=&3\Lambda _{3}/r^{4}.  \nonumber
\end{eqnarray}
The new general spacetime can be projected back into 2+1 with the result
\begin{eqnarray}
h_{ab} &=&g_{ab}-\xi _{a}^{(z)}\xi _{b}^{(z)}/\lambda _{z} \\
h_{ab}dx^{a}dx^{b} &=&F\left[ -(-m+\Lambda _{3}r^{2})dt^{2}+\frac{dr^{2}}{%
-m+\Lambda _{3}r^{2}}\right] +\frac{r^{2}}{F}d\varphi ^{2}  \nonumber
\end{eqnarray}
which can be written as
\begin{eqnarray}
h_{ab}dx^{a}dx^{b} &=&-(-m+\Lambda _{3}r^{2})dt^{2}+\frac{dr^{2}}{-m+\Lambda
_{3}r^{2}}+r^{2}d\varphi ^{2} \\
&&+(r^{4}-1)\sin ^{2}(\delta )\left[ -(-m+\Lambda _{3}r^{2})dt^{2}+\frac{%
dr^{2}}{-m+\Lambda _{3}r^{2}}+\frac{r^{2}}{1+(r^{4}-1)\sin ^{2}(\delta )}%
d\varphi ^{2}\right]   \nonumber
\end{eqnarray}
For $\delta =0$ the original 2+1 BTZ spacetime is recovered. It is also
recovered at the $r^{2}=1$ fixed point. The fluid content of the 2+1
spacetime is
\begin{eqnarray}
8\pi \rho  &=&\frac{-\Lambda _{3}}{F}+\frac{2r^{2}\sin ^{2}(\delta )}{F^{3}}%
\ \left\{ mF+(-m+\Lambda _{3}r^{2})[7\cos ^{2}(\delta )-r^{4}\sin
^{2}(\delta )]\right\}   \label{2+1content} \\
8\pi p_{r} &=&\frac{\Lambda _{3}}{F}-\frac{2r^{2}\sin ^{2}(\delta )}{F^{3}}\
\left\{ mF+2r^{4}\sin ^{2}(\delta )(-m+\Lambda _{3}r^{2})\right\}   \nonumber
\\
8\pi p_{\varphi } &=&\frac{\Lambda _{3}}{F}+\frac{2r^{2}\sin ^{2}(\delta )}{%
F^{3}}\ \left\{ 2mF+(-m+\Lambda _{3}r^{2})[5\cos ^{2}(\delta )+r^{4}\sin
^{2}(\delta )]\right\} .  \nonumber
\end{eqnarray}
The original BTZ solution described a black hole of mass $m$ surrounded by a
cosmological fluid with parameters $8\pi \rho =-\Lambda _{3},$ $8\pi
p_{r}=8\pi p_{\varphi }=\Lambda _{3}$. From Eq.(\ref{2+1content}) it is
clear that the cosmological fluid is still present but scaled by $F$, and
that in addition a new fluid has been added. For $\delta =\pi /2$, for
example, the fluid parameters are
\begin{eqnarray}
8\pi \rho  &=&(4m-3r^{2}\Lambda _{3})/r^{6}, \\
8\pi p_{r} &=&(2m-3r^{2}\Lambda _{3})/r^{6},  \nonumber \\
8\pi p_{\varphi } &=&(2m+3r^{2}\Lambda _{3})/r^{6}.  \nonumber
\end{eqnarray}
At infinity, the new solution describes an empty vacuum in contrast to the
cosmological vacuum found in the BTZ asymptotic limit. The original BTZ
solution had constant negative curvature making it locally isometric to AdS
while the new spacetime has a non-constant Ricci scalar
\begin{equation}
R=-\frac{6\Lambda _{3}}{F}+(-m+\Lambda _{3}r^{2})\frac{8r^{2}\sin
^{2}(\delta )\cos ^{2}(\delta )}{F^{3}}.
\end{equation}

\section{Discussion}

We have shown that the simple one parameter Ehlers-Geroch transform can be
applied to spacetimes with anisotropic matter content for two different
stress-energy situations. The formalism broadens the way in which anisotropy
can be introduced and studied with relation to the Killing symmetry of the
spacetime. The formalism was applied to the simple lift of the BTZ solution
and when the new 3+1 solution was projected back to 2+1, a different static
2+1 solution was obtained. It describes a 2+1 black hole with an horizon at
the same position as the original BTZ horizon but with an additional fluid
atmosphere. The asymptotic structure of the two solutions is very different.
This result suggests that it would be useful to develop the formalism in
dimensions higher than 3+1, and use the projection technique to generate and
study the resulting anisotropic 3+1 solutions. Another generalization which
could prove interesting is to broaden the fixed point structure of the
projective transform. The fixed points described by Eq.(\ref{tau-prime}) are
$\lambda =\pm 1$, as discussed above. When using the Ehlers-Geroch method
with a spacelike Killing vector, generalizing the projective transform by
placing the fixed points at $\lambda =L^{2}$ offers a better chance of
interpreting the fixed points and would broaden the applicability of the
method.

In summary, the solution generating method developed here preserves specific
Killing symmetries while creating new anisotropic solutions. These solutions
may be useful for investigating relativistic behavior at the density
extremes.

\end{document}